\documentclass[showpacs,twocolumn,prb]{revtex4-1}
\usepackage{graphicx}

\newcommand{\be}{\begin{equation}}
\newcommand{\ee}{\end{equation}}
\newcommand{\bea}{\begin{eqnarray}}
\newcommand{\eea}{\end{eqnarray}}
\newcommand{\bwt}{\begin{widetext}}
\newcommand{\ewt}{\end{widetext}}

\newcommand{\fref}[1]{Fig.~\ref{#1}}

\begin{document}

\title{Perturbation theory of the mass enhancement for a polaron coupled to acoustic phonons}

\author{Zhou Li, Carl J. Chandler, and F. Marsiglio}
\affiliation{Department of Physics, University of Alberta, Edmonton, Alberta, Canada,
T6G~2G7}

\begin{abstract}
We use both a perturbative Green's function analysis and standard perturbative quantum mechanics to calculate the decrease in energy and the effective mass for an electron interacting with acoustic phonons. The interaction is between the difference in lattice displacements for neighbouring ions, and the hopping amplitude for an electron between those two sites. The calculations are performed in one, two, and three dimensions, and comparisons are made with results from other electron-phonon models. We also compute the spectral function and quasiparticle residue, as a function of characteristic phonon frequency. There are strong indications that this model is always polaronic on one dimension, where an unusual relation between the effective mass and the quasiparticle residue is also found.
\end{abstract}

\date{\today }
\maketitle

\section{introduction}

When electrons interact strongly with phonons, the electrons acquire a polaronic character, i.e. they 
move around the lattice much more sluggishly than non-interacting electrons would, because a polarization
cloud must accompany them as they move. A measure of the strength of the coupling between the electron and
the phonons is the degree to which the ground state energy is lowered. For example, previous studies for the
Holstein model\cite{holstein59} have indicated that the decrease in energy is proportional to the bare coupling strength
($\lambda$) in strong coupling,\cite{marsiglio95} independent of the value of the phonon frequency. On the other hand, in weak coupling, while the proportionality to $\lambda$ remains, there is some dependence on phonon frequency, and in fact, the decrease in energy is greater for higher phonon frequency.\cite{marsiglio95,li10} 

A much more indicative measure of the polaronic character of an electron is the effective mass. In the Holstein model a glimpse of polaronic tendencies, even within perturbation theory, can be attained by examining the effective mass, particularly in one dimension. Usually an increasing effective mass is accompanied by a decrease in quasiparticle residue, although this is not always the case, as described below.

The Holstein model describes electrons interacting with optical phonons; the coupling is via the electron charge density, and, in this sense, the Holstein model serves as a paradigm for electron-phonon interactions just like the celebrated Hubbard model \cite{hubbard63} is the simplest description of electron-electron interactions. Many of the basic features of this model are now fairly well understood --- see Ref. [\onlinecite{fehske07,alexandrov07}] along with more recent work in Ref. [\onlinecite{li10,alvermann10}]. However, just as important is the electron interaction with acoustic phonons; typically the ionic motions couple to the electron motion, as opposed to its charge density. A very simple model to describe this kind of electron-phonon interaction within a tight-binding framework is given by

\begin{eqnarray}
H &=& -\sum_{\langle i,j \rangle} \ t_{ij} \biggl(c_{i\sigma}^\dagger c_{j\sigma} + h.c. \biggr) + \sum_i \bigl[ {p_{xi}^2 \over 2M} + {p_{yi}^2 \over 2M} \bigr]\nonumber \\
&+&  {1 \over 2} K\sum_{\langle i,j \rangle} \biggl[ \bigl(u_{xi} - u_{xj}\bigr)^2 +  \bigl(u_{yi} - u_{yj}\bigr)^2  \biggr],
\label{ham_xy}
\end{eqnarray}
where angular brackets denote nearest neighbours only, and
\begin{equation}
t_{ij}  = t -  \alpha (u_{xi} - u_{xj})\delta_{i,j\pm \hat{a}_x} - \alpha (u_{yi} - u_{yj})\delta_{i,j\pm \hat{a}_y}.
\label{hopping}
\end{equation}
This Hamiltonian has been written specifically for two dimensions, but the generalization to three dimensions (or back to one dimension) is evident from Eqs. (\ref{ham_xy}) and (\ref{hopping}). The operators and parameters are as follows: $c_{i\sigma}^\dagger$ ($c_{i\sigma}$) creates (annihilates) an electron at site $i$ with spin $\sigma$. The $x$-components for the ion momentum and displacement are given by $p_{xi}$, and displacement $u_{xi}$, respectively (similarly for the $y$-components), and the ions have mass $M$ and spring constant $K$ connecting nearest neighbours only. The electron-ion coupling is linearized in the components of the displacement, and we choose to include only longitudinal coupling.

This Hamiltonian is commonly known as the Su-Schrieffer-Heeger (SSH) model, \cite{su79,su80} because it was used for seminal work describing excitations in polyacetylene by these authors. However, it was also introduced and studied a decade earlier by Bari\u si\'c, Labb\' e, and Friedel \cite{barisic70} to describe superconductivity in transition metals, so we will refer to it as the BLF-SSH model. Much of the work done on this model is in the adiabatic approximation, i.e. the phonons are treated classically.\cite{su79,su80} This was followed by an examination of quantum fluctuations through quantum Monte Carlo and renormalization group studies,\cite{hirsch82} and these authors focused on half-filling.  They found that the lattice ordering (in one dimension) was reduced by quantum fluctuations.

Very little work has been done, however, in the quantum regime for a single electron. Capone and coworkers studied
a model similar to this one, except that they utilized optical phonons instead of acoustic ones.\cite{capone97,marchand10} This leads
to some significant differences, about which we will comment below. In the past decade Zoli has studied the BLF-SSH polaron using perturbation theory, and found, for example, a perturbative regime in one dimension where polaron effects are absent.\cite{zoli02} This result happened to agree with the conclusions of Capone et al.\cite{capone97} in the perturbative regime of the CSG model.\cite{marchand10} In this paper we focus on 2nd order perturbation theory, and find results in disagreement with Ref. [\onlinecite{zoli02}]. These results also disagree {\em qualitatively} with the results from the CSG model. That is, in one dimension, for example, perturbation theory breaks down as the characteristic phonon frequency decreases. In two dimensions there is a modest mass enhancement for all characteristic phonon frequencies, while in three dimensions the mass enhancement approaches unity in the adiabatic limit. We also note that the quasiparticle residue does not necessarily follow the trend of the inverse effective mass, as the characteristic phonon frequency varies.

This paper is organized as follows. In the following section we outline the calculation, both using perturbation theory, and using Green function techniques. For some of our work (especially in one dimension), the calculation can be done analytically, and we derive these results where applicable. In Section III we show some numerical results and compare our results with previous work and other electron-phonon models. We close in the final section with a summary.
The main conclusion is that, as far as one can tell from weak coupling perturbation theory, the BLF-SSH model has a stronger tendency to form a polaronic state than is the case with the Holstein model. In one dimension this is most evident in the effective mass, and not at all evident in the quasiparticle residue.

\section{perturbation theory}

\subsection{Hamiltonian}

The Hamiltonian Eq. (\ref{ham_xy}), Fourier-transformed to wavevector space, and utilizing phonon creation and annihilation operators, is written (again in 2D),
\begin{eqnarray}
H &=& \sum_{k\sigma} \epsilon_{k\sigma} c_{k\sigma}^\dagger c_{k\sigma} \nonumber \\
&+& \sum_q \hbar \omega(q) \bigl[ a_{xq}^\dagger a_{xq} + a_{yq}^\dagger a_{yq} \bigr] \nonumber \\
&+& \sum_{{k k^\prime} \atop \sigma} g_{x}(k,k^\prime) \bigl[ a_{x k-k^\prime} + a_{x -(k-k^\prime)}^\dagger \bigr] c_{k\sigma}^\dagger c_{k^\prime \sigma} \nonumber \\
&+& \sum_{{k k^\prime} \atop \sigma} g_{y}(k,k^\prime) \bigl[ a_{y k-k^\prime} + a_{y -(k-k^\prime)}^\dagger \bigr] c_{k\sigma}^\dagger c_{k^\prime \sigma}.
\label{ham_q}
\end{eqnarray}
Here, 
\begin{equation}
\epsilon_k \equiv \epsilon(k_x,k_y) = -2t [\cos{(k_x)} + \cos{(k_y)}]
\label{electron_dispersion}
\end{equation}
is the dispersion relation for non-interacting electrons with nearest neighbour hopping, and
\begin{equation}
\omega(q) \equiv \omega_0 \sqrt{ \sin^2{(q_x/2)} + \sin^2{(q_y/2)}}
\label{phonon_dispersion}
\end{equation}
is the phonon dispersion for acoustic phonons with nearest neighbour spring constants $K$, and $\omega_0 \equiv \sqrt{4K/M}$ is the characteristic phonon frequency. The phonon creation and annihilation operators are given by
$a_{x q}^\dagger$ and $a_{x q}$, respectively, and similarly for those in the y-direction. The coupling "constants"
are given by
\begin{equation}
g_x(k,k^\prime) \equiv i \alpha \sqrt{\frac{2}{MN\omega(k-k^\prime)}} \biggl[\sin{(k^\prime_x)} - \sin{(k_x)} \biggr],
\label{gkkprime}
\end{equation}
with a similar expression for the $y$ direction, and $M$ is the mass of the ion and $N$ is the number of lattice sites.

\subsection{Green's function analysis}

Carrying out a Green's function analysis using the free electron and phonon parts of the Hamitonian as the unperturbed part, gives, for the self energy of a single electron to lowest (2nd) order in the coupling $\alpha$,
\begin{eqnarray}
& & \Sigma(k,\omega+ i\delta)  =  \nonumber \\
\label{selfenergy}
& & -\sum_{k^\prime}\bigl[ |g_x(k,k^\prime)|^2 + |g_y(k,k^\prime)|^2 \bigr] G_0(k^\prime,\omega + i\delta - \omega(k-k^\prime)), \nonumber \\
\end{eqnarray}
where $G_0(k,\omega+i\delta) \equiv \bigl[ \omega + i\delta - \epsilon_k \bigr]^{-1}$ is the non-interacting electron retarded propagator. 

One way to determine the effect of interactions on the electron dispersion is to compute the renormalized energy for the ground state (here, $k_x = k_y = 0$), and the effective mass. The effective mass has long been used as the primary
indicator for polaronic behaviour \cite{fehske07,alexandrov07}, and though within 2nd order perturbation we can only get an indication of this crossover, we use it here nonetheless. The renormalized energy is given by the solution for the pole location in the interacting electron Green's function, $G(k,\omega+i\delta) \equiv \bigl[ \omega + i\delta - \epsilon_k - \Sigma(k,\omega + i\delta)\bigr]^{-1}$,
\begin{equation}
E_k = \epsilon_k + {\rm Re}\Sigma(k,E_k).
\label{renormalized_energy}
\end{equation}
To determine the effective mass, defined by the expectation that $E_k \equiv \hbar^2 k^2/(2m^\ast)$, we take two derivatives\cite{remark1} of Eq. (\ref{renormalized_energy}), and, using the fact that $(dE_k/dk)|_{k=0} = 0$, we obtain
\begin{eqnarray}
\frac{m^\ast}{m} &= &{1 - {\partial \Sigma(k,\omega) \over \partial \omega}|_{\omega = E_k} \over
1 + \frac{1}{2t} {\partial^2 \Sigma(k,\omega) \over \partial k^2}|_{\omega = E_k} }\nonumber \\
&=&
1 - {\partial \Sigma(k,\omega) \over \partial \omega}|_{\omega = E_k} - \frac{1}{2t} {\partial^2 \Sigma(k,\omega) \over \partial k^2}|_{\omega = E_k}.
\label{effective_mass}
\end{eqnarray}
Here we have used the fact that the band mass given by the electron dispersion in Eq. (\ref{electron_dispersion})
is $m = 1/(2t)$. Note that it is common (and advisable) to replace the substitutions for $\omega$ required in Eq. (\ref{effective_mass}) with $\epsilon_k$, rather than with $E_k$. This is due to the fact that the former substitution keeps the evaluation for every term at $O(\alpha^2)$, whereas the latter substitution includes some (inconsistently) higher order contributions. The former substitution is known as Rayleigh-Schrodinger perturbation theory while the latter is known as Brillouin-Wigner perturbation theory.\cite{mahan00} This means that  we will use the following equation,
\begin{equation}
\frac{m^\ast}{m} = 1 - {\partial \Sigma(k,\omega) \over \partial \omega}|_{\omega = \epsilon_k} - \frac{1}{2t} {\partial^2 \Sigma(k,\omega) \over \partial k^2}|_{\omega = \epsilon_k},
\label{effective_mass_new}
\end{equation}
to define the effective mass.

In contrast the quasiparticle residue is defined as the weight that remains in the $\delta$-function-like portion of the spectral weight. The spectral weight is defined as
\begin{eqnarray}
A(k,\omega) &\equiv & -{1 \over \pi} {\rm Im} G(k,\omega + i\delta) \nonumber \\
& = & - {1 \over \pi} {\rm Im} { 1 \over \omega + i\delta - \epsilon_k - \Sigma(k,\omega + i\delta)}.
\label{spectral}
\end{eqnarray}
For a given momentum, as the energy of the pole given by Eq. (\ref{renormalized_energy}) is approached, the imaginary part of the self energy tends towards zero; this produces a $\delta$-function contribution in Eq. (\ref{spectral}) , at the pole energy, but with weight $z_k$ defined by
\begin{equation}
z_k = {1 \over 1 - {\partial \Sigma(k,\omega) \over \partial \omega}|_{\omega = E_k} }.
\label{residue}
\end{equation}
The relationship amongst these various quantities --- effective mass in Eq. (\ref{effective_mass}), effective mass in Eq. (\ref{effective_mass_new}), and quasiparticle residue in Eq. (\ref{residue}) --- is discussed further in the Appendix.

\subsection{Standard perturbation theory}

Eq. (\ref{effective_mass}) requires a numerical evaluation of Eq. (\ref{selfenergy}), and then the required derivatives can be (numerically) determined. Because the positions of the singularities in Eq. (\ref{selfenergy}) are
difficult to determine in advance, it is customary to introduce a small (numerical) imaginary part corresponding to the infinitesimal $\delta$, and then the numerical integration is more stable. This trick remains problematic, as we discuss further below. Alternatively, we can simply perform a 2nd order perturbation theory expansion, as outlined in every undergraduate quantum mechanics textbook. The result is
\begin{equation}
E^{(2)}_k = \frac{2\alpha^2}{M} \frac{1}{N} \sum_{k^\prime}  { \bigl( \sin{k_x^\prime} - \sin{k_x} \bigr)^2 + \bigl( \sin{k_y^\prime} - \sin{k_y} \bigr)^2 \over \omega(k-k^\prime) \ \ \  [\epsilon_k - \epsilon_{k^\prime} - \omega(k-k^\prime)]},
\label{rs_pert}
\end{equation}
where we remember that the first order (in $\alpha$) contribution is of course zero, and the superscript $(2)$ indicates the 2nd order contribution. Comparison with Eq. (\ref{selfenergy}) shows that this corresponds to Rayleigh-Schrodinger perturbation theory with the self energy, evaluated at $\omega = \epsilon_k$ corresponding to the 2nd order energy correction. Eq. (\ref{rs_pert}) can be evaluated numerically, and then two derivatives with respect to $k$ are required. However, the same numerical problems mentioned above will arise; fortunately, at least in one dimension, Eq. (\ref{rs_pert}) can be evaluated analytically, whereas we were unable to do the same with Eq. (\ref{selfenergy}).

\section{results and discussion}

\subsection{Analytical results in 1D}

The result of an analytical evaluation\cite{remark2} of Eq. (\ref{rs_pert}) is, in one dimension,
\begin{equation}
E^{(2)}(k) = -\frac{32t}{\pi} \lambda_{\rm BLF} \tilde{\omega}_0 \biggl\{-2 \cos{k} + \pi \tilde{\omega}_0 + C_k(\tilde{\omega}_0) \biggr\},
\label{2nd_order_anal}
\end{equation}
where  $\tilde{\omega}_0 \equiv {\omega}_0/(4t)$, and a dimensionless coupling parameter $\lambda_{\rm BLF}$ is
defined, in analogy to the dimensionless coupling parameter defined in the Holstein model, as
\begin{equation}
\lambda_{\rm BLF} \equiv \frac{\alpha^2}{M \omega_0^2} \frac{1}{W},
\label{lam_blf}
\end{equation}
where here the bandwidth $W = 4t$ for one dimension. Note that this coupling parameter has nothing to do physically with the coupling parameter defined in the Holstein model, so we will treat them as completely independent.\cite{remark3} The function $C_k(\tilde{\omega}_0)$ must be evaluated separately in the two regimes:
\begin{equation}
C_k(\tilde{\omega}_0)  = 2 \sqrt{\tilde{\omega}_0^2 - 1} \biggl( h(k) + h(-k) - 2h(\pi/2) \biggr), \ \ \ \ \tilde{\omega}_0 > 1,
\label{ck_greater}
\end{equation}
where 
\begin{equation}
h(k) = {\rm tan}^{-1} \biggl( {\tilde{\omega}_0 {\rm tan} {k \over 2} + 1 \over \sqrt{\tilde{\omega}_0^2 - 1}} \biggr)
\label{hp}
\end{equation}
and
\begin{equation}
C_k(\tilde{\omega}_0)  = \sqrt{1 - \tilde{\omega}_0^2} \biggl( s(k) + s(-k) - 2s(\pi/2) \biggr), \ \ \ \ \tilde{\omega}_0 < 1,
\label{ck_lesser}
\end{equation}
where 
\begin{equation}
s(k) = {\rm log}\Biggl( {\tilde{\omega}_0  {\rm tan} {k \over 2} + 1 + \sqrt{1 - \tilde{\omega}_0^2} \over
 \tilde{\omega}_0  {\rm tan} {k \over 2} + 1 - \sqrt{1 - \tilde{\omega}_0^2}  }\Biggr).
\label{sp}
\end{equation}
Eq. (\ref{2nd_order_anal}) is readily evaluated at $k=0$ to determine the ground state energy. Evaluating the second derivative with respect to wave vector $k$ is equally straightforward, and determination at $k=0$ yields the rather simple result for the effective mass,
\begin{equation}
\frac{m^\ast}{m} = 1 + \frac{32}{\pi} \frac{\lambda_{\rm BLF}}{\tilde{\omega}_0},
\label{eff_mass_anal}
\end{equation}
valid for all values of $\tilde{\omega}_0$.\cite{remark_k}

\subsection{ Comparison with other models}

An analytical result is readily available for the Holstein model; there, the ground state energy (in 1D) was
given by\cite{marsiglio95}
\begin{equation}
E_{\rm H} = -2t \biggl( 1 + \lambda_{\rm H} \sqrt{\tilde{\omega}_E \over \tilde{\omega}_E + 1} \biggr),
\label{energy_hol}
\end{equation}
where $\tilde{\omega}_E \equiv \omega_E/(4t)$ is the Einstein phonon frequency normalized to the bandwidth,
and, as explained earlier, the dimensionless coupling constant $\lambda_{\rm H}$ cannot be compared directly
to the corresponding quantity for the BLF-SSH model.
The effective mass is given by
\begin{equation}
\biggl({m^\ast \over m}\biggr)_{\rm H} = 1 + {\lambda_{\rm H} \over 4 \sqrt{\tilde{\omega}_E}} { 1 + 2\tilde{\omega}_E \over 
\bigl( 1 + \tilde{\omega}_E \bigr)^{3/2}}.
\label{mass_hol}
\end{equation}
In both cases, as the characteristic phonon frequency approaches zero (adiabatic limit) the ground state energy approaches the non-interacting value; however, the effective mass diverges in this same limit. So, while the first statement would appear to justify perturbation theory in this limit, the second statement clearly indicates a breakdown in the adiabatic limit. It is known in both cases that the adiabatic approximation leads to a polaron-like solution for all
coupling constants,\cite{kabanov93,chandler10} and clearly these two observations are consistent with one another. In fact, the divergence is stronger in the BLF-SSH model, and goes beyond the inverse square-root behaviour observed for the Holstein model and attributed to the diverging electron density of states in one dimension;\cite{capone97} this indicates that the BLF-SSH model, at least in the adiabatic limit in one dimension, has a stronger tendency for polaron formation than the Holstein model.

Interestingly, in the model studied by Capone et al.\cite{capone97}, where optical phonons were used, the opposite behaviour was obtained; they found that the effective mass ratio approached unity as the characteristic phonon energy approached zero.\cite{remark4} In the opposite limit Capone et al.\cite{capone97} found an effective mass ratio that did not approach unity as the characteristic phonon frequency increased (anti-adiabatic limit). In the BLF-SSH model, however, this ratio does approach unity as the phonon frequency increases beyond the electron bandwidth, in one dimension, in agreement with the Holstein result in all dimensions. As we will see below, however, in the BLF-SSH model in two and three dimensions the effective mass ratio remains above unity in the anti-adiabatic limit. This is not surprising, since here the interaction modulates the hopping, and we expect a non-zero correction in this 
limit.\cite{remark4} In the adiabatic limit, the BLF-SSH mass ratio approaches a constant value in two dimensions, and falls to unity in three dimensions, both in agreement with the behaviour in the Holstein model.

Our results disagree with those of Zoli\cite{zoli02} for reasons that are not entirely clear. We have utilized both the straightforward perturbation theory method (analytically and numerically), and the Green's function formalism (numerically). In the latter case we required a numerically small imaginary part for the frequency significantly smaller than the value quoted in Ref. (\onlinecite{zoli02}) (we used $\delta = 10^{-9}$ whereas he used $\delta = 10^{-4}$. However, as is clear from our analytical result, Eq. (\ref{eff_mass_anal}), our effective mass diverges at low phonon frequency, and decreases monotonically to unity as the phonon frequency increases. The result in Ref. (\onlinecite{zoli02}) peaks sharply near $\tilde{\omega}_0 \approx 1$, and, as noted above, decreases to unity at low phonon frequency.

\begin{figure}[tp]
\begin{center}
\includegraphics[height=3.5in,width=3.5in,angle=0]{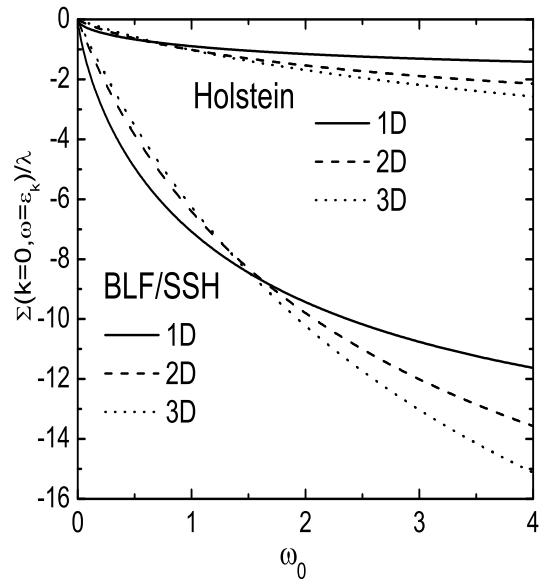}
\end{center}
\caption{ Electron self energy for the ground state ($k=0$), normalized to $\lambda$ (or $\lambda_H$) vs. characteristic phonon frequency $\omega_0$ (this is $\omega_E$ for the Holstein model), for both the BLF-SSH and Holstein models, in one, two, and three dimensions, as indicated. Alternatively, the ordinate is simply the second order (in $g$) correction to the ground state energy within Rayleigh-Schrodinger perturbation theory. In all cases the magnitude of the correction increases with increasing $\omega_0$. At low $\omega_0$ the magnitudes of the the results are ordered 3D, 2D, 1D (lowest to highest) whereas at high frequency the ordering is just the opposite.  All six cases have non-zero limiting values as $\omega_0 \rightarrow \infty$, given in Table 1.}
\label{fig1}
\end{figure}

\subsection{ Numerical results}

In \fref{fig1} we plot the reduction in the ground state energy due to the second order correction (for the BLF-SSH model, this is given by Eq. (\ref{rs_pert})), normalized
to $\lambda$ (or $\lambda_H$). This is also written as $\Sigma(k=0,\omega = \epsilon_k)/\lambda$, where the self energy is given by the expression in Eq. (\ref{selfenergy}). Also plotted for comparison are the corresponding quantities for the Holstein model.
Note that both models share a few features in common: (i) they both go to zero as the characteristic phonon energy decreases to zero, regardless of the dimensionality, (ii) they all approach a non-zero negative (and finite) value as
the characteristic phonon frequency grows, and (iii) they cross one another in strength as a function of dimensionality as $\omega_0$ increases, i.e. at low phonon frequencies the self energy has the highest magnitude for one dimension, whereas for high phonon frequency  the highest magnitude is achieved in both models for three dimensional systems.
Also note that the BLF-SSH results are well separated from Holstein results. In particular, there appears to be more 'bang for the buck' with the BLF-SSH model, i.e. for a given value of $\lambda$ and the same characteristic phonon frequency, the energy reduction is almost an order of magnitude higher for the BLF-SSH model as compared with the Holstein model. Again, we remind the reader that the value of $\lambda$ in the Holstein model has nothing to do with the value of $\lambda$ in the BLF-SSH model, so this comparison is unwarranted.

For this reason we will use the value for the self energy, in weak coupling, as the phonon frequency increases to infinity, as the energy scale that provides a measure of the energy lowering expected for a given model and a given dimensionality. These numbers, mostly determined analytically, are provided in Table I.


\begin{table}[ht] 
\caption{$\lim_{\omega_0 \rightarrow \infty} \Sigma(k=0,\omega =\epsilon_k)/(\lambda t)$} 
\centering 
\begin{tabular}{ccc} 
\hline\hline 
Dim. & BLF-SSH & Holstein  \\ [0.5ex] 
\hline 
1D & -16 & -2 \\ 
2D & -23.3 & -4  \\ 
3D & -30.2 & -6  \\ [1ex] 
\hline 
\end{tabular} 
\label{table:table1} 
\end{table}

\begin{figure}[tp]
\begin{center}
\includegraphics[height=3.5in,width=3.5in]{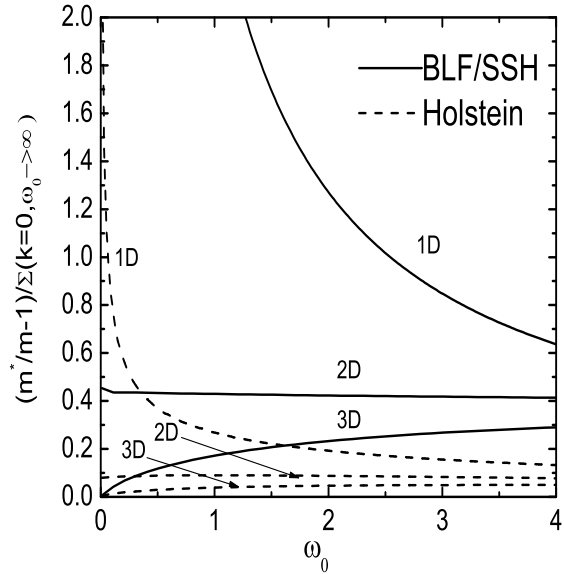}
\end{center}
\caption{ The electron effective mass, normalized to the 2nd order correction to the energy for the anti-adiabatic limit, vs. characteristic phonon frequency, $\omega_0$, for both the BLF-SSH and Holstein models, in one, two, and three dimensions, as indicated. In 1D the effective mass diverges for both models, though the divergence is stronger for the BLF-SSH model, as indicated by Eq. (\protect\ref{eff_mass_anal}). In 2D the effective mass approaches a constant as $\omega_0 \rightarrow 0$ for both models, while in 3D the effective mass ratio approaches unity in the same limit. At the opposite extreme, both 1D results give $m^\ast/m \rightarrow 1$ as $\omega_0 \rightarrow \infty$, while in both 2D and 3D the effective mass remains above unity in this limit. Note that in all three dimensions, for a given reduction in energy as given by the 2nd order correction to the energy, the BLF-SSH model results in significantly higher effective masses.}
\label{fig2}
\end{figure}

In  \fref{fig2} we plot the effective mass ratio (minus unity), normalized to the self energy evaluated for infinite characteristic phonon frequency. This normalization is important to divide out enhancements that are solely
due to definitions. Moreover, in this way, we are determining the mass enhancement for a given 'coupling strength', where this strength is now a measure of the energy lowering caused by a certain amount of coupling to phonons, regardless of the origin of that coupling. This plot now makes clear that the BLF-SSH model, within weak coupling perturbation theory, has more 'polaronic' tendency than the Holstein model. Note in particular that the divergence (in 1D) at low characteristic phonon frequency is much stronger for the BLF-SSH model, as Eq. (\ref{eff_mass_anal}) already indicated. Thus, as discussed above, we anticipate that in the adiabatic approximation, in 1D, the system will always be polaronic, regardless of the coupling strength, in agreement with the result of the Holstein model,\cite{kabanov93} and in disagreement with the result from the hybrid model defined in Ref. \onlinecite{capone97}.

\begin{figure}[tp]
\begin{center}
\includegraphics[height=3.5in,width=3.5in]{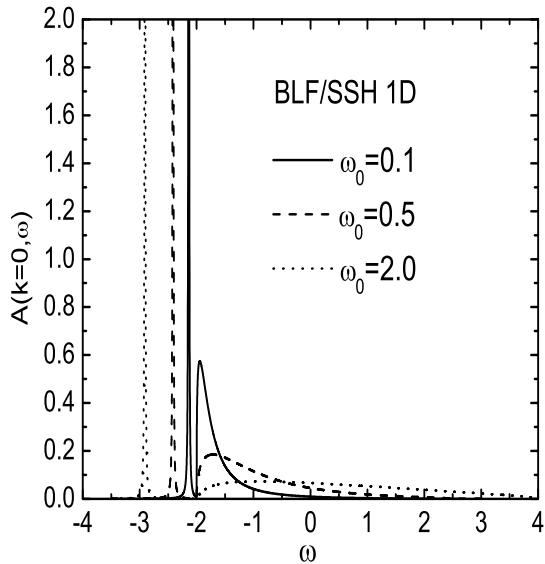}
\end{center}
\caption{Spectral function for the BLF-SSH model, for $\lambda = 0.2$ for three different characteristic phonon frequencies, as a function of frequency. All three spectra are similar as one would find for the Holstein model, and consist of quasiparticle peak with weight $z_0 = 0.766, 0.727, 0.724$, for $\omega_0/t = 0.1, 0.5, 2.0$, respectively, followed by an incoherent piece.}
\label{fig3}
\end{figure}

Otherwise, the behaviour of the effective mass in the two models is very similar, as a function of characteristic phonon frequency, for the various dimensions shown. The effective mass can be made arbitrarily close to unity, for any non-zero phonon frequency, for sufficiently weak coupling. Preliminary numerical calculations indicate a free electron-like to polaron crossover,\cite{li10b} similar to what was found for the Holstein model.

\subsection{ Spectral function}

\begin{figure}[tp]
\begin{center}
\includegraphics[height=3.0in,width=3.0in,angle=0]{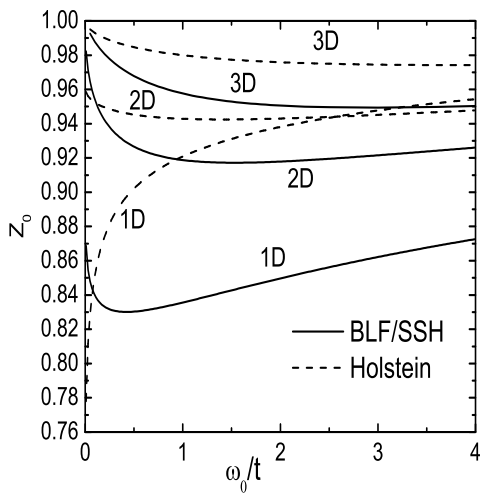}
\end{center}
\caption{ Quasiparticle residue, $z_0$ vs. $\omega_0/t$ for both the BLF-SSH and Holstein models. in all three dimensions. Note that while the result for the Holstein model tends to be inversely proportional to the effective mass, this is not the case for the BLF-SSH model at low phonon frequency, and in 1D and 2D. In one dimension in particular, the effective mass diverges, while $z_0$ also turns upward.}
\label{fig4}
\end{figure}

It is interesting to examine the spectral function, defined by Eq. (\ref{spectral}) (see also the discussion in the Appendix).
For simplicity we show the result in one dimension, in \fref{fig3}, for the ground state ($k=0$) as a function of frequency.

The results for two or three dimensions do not differ in any significant way from these results. 
The results for three different characteristic phonon frequencies are shown. In each case a quasiparticle 
$\delta$-function is present (here artificially broadened so as to be visible), followed by an incoherent piece; the incoherent part has energies ranging approximately from $-2t < \omega < +2t + \omega_0$. The quasiparticle residue, $z_0$ must be determined numerically, and is given in the figure caption for each of the cases considered (see also 
\fref{fig4}). We have verified that the remaining weight (the spectral functions each have weight unity) is present in the incoherent part.
\begin{figure}[tp]
\begin{center}
\includegraphics[height=3.0in,width=3.0in,angle=0]{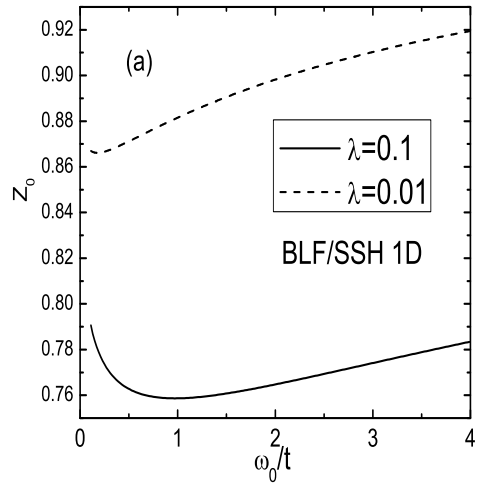}
\includegraphics[height=3.0in,width=3.0in,angle=0]{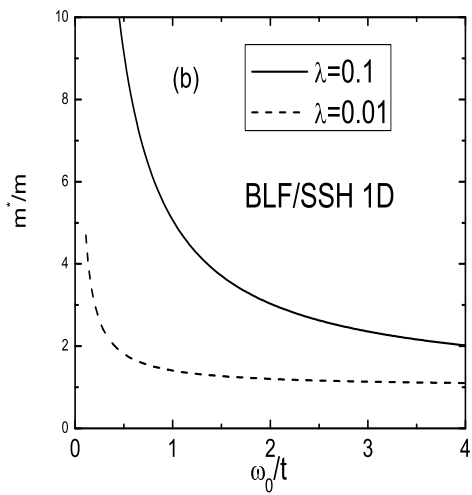}
\end{center}
\caption{ Comparison of the quasiparticle residue (upper panel) with the electron effective mass (lower panel) as a function of $\omega_0/t$, for the BLF-SSH model in one dimension. The behaviour noted in \fref{fig4} is clear here. Moreover, note the scales; while the effective mass ratio is very large ($\approx 4$) for $\lambda = 0.01$ and small values of $\omega_0/t$, the quasiparticle residue remains within 15\% of unity.}
\label{fig5}
\end{figure}
The result shown is not too different from what is found in the Holstein model; the singularities from the 1D electron density of states are now smeared out in the incoherent piece, as a result of the coupling and phonon energy having some frequency dependence. We show in \fref{fig4}, as a function of $\omega_0$, the quasiparticle residue for both the Holstein and BLF-SSH models. The Holstein results tend to follow the inverse of the result for the inverse effective mass; this is as expected. This is not the case with the BLF-SSH, but for more subtle reasons than the fact that the self energy is now momentum dependent. The more important effect, which shows up in both 1D and 2D results, is that the quasiparticle weight requires an evaluation of the frequency derivative of the self energy at the energy of the pole, whereas the effective mass in Rayleigh-Schrodinger perturbation theory requires the same derivative at the {\em non-interacting} ground state energy. Most noteworthy is that the quasiparticle residue shows a clear upturn at low characteristic phonon frequencies, while the inverse effective mass clearly approaches zero (see \fref{fig2}) as this characteristic frequency is taken to zero. 

To see this more clearly we show in \fref{fig5} a comparison of the residue (upper panel) vs. effective mass (lower panel), as a function of $\omega_0$, for two (weak) strengths of electron phonon coupling. At high phonon frequency, as the former decreases, the latter increases with decreasing phonon frequency, but at low phonon frequency, the two properties no longer behave in inverse fashion with respect to one another.

\section{Summary}

The BLF-SSH model appears to have very strong polaronic tendencies, stronger than those of, say, the Holstein model, especially in one dimension. This conclusion is based on the 2nd order perturbative calculation performed in this paper, but also has corroborative evidence from calculations in the strong coupling regime. In one dimension we have been able to obtain an analytical solution for the ground state energy and the effective mass. The conclusion concerning polaronic behaviour is an important one, as much of what we know about polarons arises from Holstein-like models.\cite{remark5} In particular, for a coupling strength that leads to a fixed amount of energy lowering (in 2nd order), the effective mass can become an order of magnitude larger than the bare mass, a clear indicator that perturbation theory breaks down. This occurs in the BLF-SSH model at much weaker coupling than in the Holstein model. We have also noted that the relationship between effective mass and quasiparticle residue breaks down in one and two dimensions for the BLF-SSH model, not because of the momentum dependence in the self energy, but because the two properties involve evaluation of the frequency derivative of the self energy at different energies. Future work will address the strong coupling regime.

\begin{acknowledgments}

This work was supported in part by the Natural Sciences and Engineering
Research Council of Canada (NSERC), by ICORE (Alberta), by Alberta Ingenuity, and by the Canadian
Institute for Advanced Research (CIfAR). CC was supported by an NSERC USRA and ZL was supported by an Alberta Ingenuity Fellowship.

\end{acknowledgments}

\appendix
\section{ Perturbation Theory}

It is sometimes stated that for a momentum-independent self  energy, the quasiparticle residue is equal to the inverse of the effective mass. This follows simply by comparing Eqs. (\ref{effective_mass}) and (\ref{residue}). On the other hand, we have argued that Eq. (\ref{effective_mass_new}) is more appropriate for the effective mass, in which case this statement appears not to be true. A resolution of this difficulty is straightforward for the Holstein model, which we outline below, but, interestingly, not possible for the BLF-SSH model, at least in one dimension. The essential difference appears to be that in the Holstein model the (phonon) excitations are gapped, whereas they are not in the BLF-SSH model because of the low-lying acoustic modes at small momentum transfer. In this appendix we focus attention on one dimension, where some subtleties arise.

For the Holstein model the computation of the self energy in weak coupling is straightforward.\cite{marsiglio95} We obtain
\begin{equation}
\Sigma_H(\omega) = {2t \omega_E \lambda_H {\rm sgn}(\omega - \omega_E) \over
\sqrt{(\omega- \omega_E)^2 - (2t)^2}}.
\label{a_holstein_self}
\end{equation}
The location of the quasiparticle pole at zero momentum (ground state) is then given by
\begin{equation}
\omega + 2t = -{2t \omega_E \lambda_H  \over
\sqrt{(\omega- \omega_E)^2 - (2t)^2}},
\label{a_holstein_pole}
\end{equation}
which can readily be determined numerically. Denoting the solution by writing $\omega \equiv -2t - E_b$ (so $E_b$ is the 'binding' energy below the bottom of the band),  we can then use this in the spectral function, Eq. (\ref{spectral}), to determine the residue $z_0$ in the quasiparticle peak at $\omega = -2t - E_b$:
\begin{equation}
A(k=0,\omega) = z_0 \delta(\omega + 2t + E_b) + {\rm incoherent \  part}.
\label{a_spec}
\end{equation}
Straightforward calculation gives
\begin{equation}
z_0 = 1/\biggl(1 + { 2\lambda_H \tilde{\omega}_E \bigl[1 + 2\tilde{\omega}_E + 2\tilde{E}_b \bigr] \over \bigl[ (1 + 2\tilde{\omega}_E + 2\tilde{E}_b )^2 - 1\bigr]^{3/2}}    \biggr),
\label{a_z_0}
\end{equation}
which is {\em not} in agreement with the inverse of Eq. (\ref{mass_hol}), except when $\lambda_H$ is truly very small.
Here $\tilde{E}_b \equiv E_b/(4t)$.

In particular, for arbitrarily small $\lambda_H$, $\partial \Sigma(\omega)/\partial \omega |_{\omega = -2t}$, which is
used in Eq. (\ref{mass_hol}), diverges as $\omega_E \rightarrow 0$, leading to a divergent effective mass (and therefore associated residue of zero).
On the other hand, from Eq. (\ref{a_holstein_pole}) one readily sees
\begin{equation}
\lim_{\omega_E \rightarrow 0} {E}_b = t\bigl(\lambda \omega_E/t \bigr)^{2/3},
\label{a_energy}
\end{equation}
from which Eq. (\ref{a_z_0}) yields the result
\begin{equation}
\lim_{\omega_E \rightarrow 0} z_0 = 2/3,
\label{a_z_00}
\end{equation}
surprisingly a universal number. The actual weight in the quasiparticle peak of the spectral function given by Eq. (\ref{spectral}) for any given (even very small) value of $\lambda_H$ actually tracks Eq. (\ref{a_z_0}), and not the inverse of Eq. (\ref{mass_hol}).

Interestingly, for the Holstein model, one can take a different tact towards calculating the spectral function: using perturbation theory to compute the perturbed wave function, which is then inserted into the calculation for the matrix elements required in the definition of the spectral function,\cite{remark5} one obtains
\begin{eqnarray}
A_{\rm pert}(k=0,\omega&) &= z_0^{\rm pert} \delta (\omega +2t +{\lambda_H \omega_E \over \sqrt{(1+2\tilde{\omega}_E)^2) - 1}}) \nonumber \\
& + &{1 \over \pi} {2t \omega_E \lambda_H \over (\omega+2t)^2}{\theta(2t - |\omega - \omega_E|) \over \sqrt{(2t)^2 - (\omega - \omega_E)^2}}.
\label{a_spec2}
\end{eqnarray}
Note that there is no difficulty in integrating over this function, as the divergence in the denominator ($1/(\omega + 2t)^2$) is not within (or bordering) the range of frequency given by the Heaviside function restriction in the numerator. This is due to the finite phonon frequency, $\omega_E$.
From this expression fulfillment of the sum rule determines that
\begin{equation}
z_0^{\rm pert} = 1/\biggl(1 + { 2\lambda_H \tilde{\omega}_E \bigl[1 + 2\tilde{\omega}_E  \bigr] \over \bigl[ (1 + 2\tilde{\omega}_E)^2 - 1\bigr]^{3/2}}    \biggr),
\label{a_z_0_pert}
\end{equation}
which {\em is in agreement} with the inverse of Eq. (\ref{mass_hol}). The message is that, as long as we use the expression given by Eq. (\ref{spectral}) for the spectral function, the area under the quasiparticle peak will correspond to Eq. (\ref{a_z_0}), which is {\em not} the inverse of the effective mass, even if the self energy is independent of momentum.

In the BLF-SSH model, the self energy is evaluated numerically through Eq. (\ref{selfenergy}). An attempt to follow the procedure just outlined, which leads to Eqs. (\ref{a_spec2}) and (\ref{a_z_0_pert}) for this model fails; this is because the minimum phonon frequency is zero, so the restriction corresponding to the Heaviside function in Eq. (\ref{a_spec2}) yields $-2t < \omega < 2t + \omega_0$; this in turn makes the divergence at $\omega = -2t$  non-integrable. One can only (in 1D) define the spectral function through Eq. (\ref{spectral}), in which case the inverse of the effective mass differs from the quasiparticle pole for two reasons: the usual reason that the explicit momentum dependence now plays a role (see Eq. (\ref{effective_mass_new})), and, in addition, the derivative of the self energy with respect to frequency is evaluated at 
$\omega = -2t$ for the effective mass, whereas it is evaluated at the frequency corresponding to the pole for the quasiparticle residue.


\begin{thebibliography}{99}
\bibitem{remark1} The effective mass will in general be anisotropic, so the direction in which these derivatives are to be taken would have to be specified; we nonetheless retain this simple notation and avoid some cumbersome indices.
\bibitem{remark2} One needs to use the identities $\sin{a} - \sin{b} = 2 \sin{[(a-b)/2]} \cos{[(a+b)/2]}$ and $\cos{a} - \cos{b} = -2 \sin{[(a-b)/2]} \sin{[(a+b)/2]}$, and then the integral is straightforward.
\bibitem{remark3} However, since in this model the coupling is to the motion of the electron, for a given 'strength' of coupling as given by $\alpha$ in this model, we expect the overall contribution here to grow as the number of nearest neighbours increases, whereas the same is not true for the Holstein model.
\bibitem{remark4} As explained in Ref. (\onlinecite{capone97}), we utilize limiting procedures for the characteristic phonon frequency while keeping the effective spring constant fixed. This means, for example, that as the phonon frequency approaches zero the ion mass increases to keep the product $M\omega_0^2$ constant.
\bibitem{remark5} We include Frohlich models in this category, as the lattice displacement couples to the charge density, as occurs in the Holstein model; the interaction is simply long range.
\end{thebibliography}
\end{document}